\documentstyle[12pt,aasms4]{article}

\begin{document}

\title{On the Structure and Morphology of the `Diffuse Ionized Medium'
in Star-Forming Galaxies}
 
\author{Jing Wang$^{1,3}$, Timothy M. Heckman$^{1,3,4}$, and Matthew D.
Lehnert$^{2,3}$}
 
\parindent=0em
\vspace{5cm}
 
1. Department of Physics and Astronomy, The Johns Hopkins University,
Baltimore, MD 21218
 
2. Leiden Observatory, Postbus 9513, 2300RA, Leiden, The Netherlands
 
3. Visiting observers at the Kitt Peak National Observatory of the
National Optical Astronomy Observatories, operated by AURA under
contract with the National Science Foundation.
 
4. Adjunct Astronomer, Space Telescope Science Institute.
 
\parindent=2em
 
\begin{abstract}

  Deep H$\alpha$ images of a sample of nearby late-type spiral
galaxies have been analyzed to characterize the morphology and energetic
significance of the
``Diffuse Ionized Medium'' (DIM). 
We find that the DIM properties can be reasonably unified
as a function of {\it relative} surface brightness ($\Sigma / \bar{\Sigma}$,
where $\bar{\Sigma}$ is the mean H$\alpha$ surface brightness within regions
lying above a fixed very faint isophotal level). We measured the images 
down to this common isophotal limit and constructed
a fundamental dimensionless surface-brightness distribution function
that describes the dependence of the area 
(normalized to total area) occupied by gas with a given relative surface
brightness ($\Sigma / \bar{\Sigma}$). 
This function determines both
the flux and area contribution by the DIM to global values. The
function is found to be almost the same at high surface brightness 
($\Sigma / \bar{\Sigma} \gtrsim 1$) and less similar for 
$\Sigma / \bar{\Sigma} \lesssim 1$. We show the universal distribution function
at high surface brightness can be understood as 
a consequence of the general properties of HII regions
including their
H$\alpha$ luminosity function and exponential radial brightness
profiles. We suggest that {\it relative} surface brightness (rather than
an absolute value) is a more physically-meaningful criterion to discriminate
the DIM from HII regions. The use of the dimensionless distribution function
to quantify the DIM is consistent with the fundamentally morphological
definition of the DIM as being `Diffuse'.
The difference in the distribution function from galaxy to galaxy
at low surface brightness quantifies the different prominence
of the DIM in the galaxies. This variation is found to be consistent with
results from other complementary ways of determining the
DIM's global importance. The variation of the DIM among the galaxies
that is 
indicated by
the distribution function is small enough to guarantee that the fractional
contribution of the DIM to the global H$\alpha$ luminosity in the
galaxies is fairly
constant, as has been observed. The {\it continuous} transition from 
HII regions to the DIM in the distribution function suggests that
the ionizing energy for the DIM mainly comes from HII regions, consistent
with the ``leaky HII regions model''.

\end{abstract}
 
 
\section{INTRODUCTION}
 
The interaction/feedback between massive stars and the interstellar medium
has been a key problem in understanding star formation
and the evolution of galaxies. The physical state, dynamics and
composition of the multiphase ISM are greatly influenced
by the energy and material deposited by stars. 
The most recently discovered
major component of the ISM is the widespread diffuse ionized medium 
(DIM). This gas, first discovered as
the `Reynolds Layer' in our Galaxy (see, e.g. Reynolds 1990 for a review),
and then in other late-type spiral (e.g. Ferguson et al. 1996;
Hoopes et al. 1996;
Wang, Heckman \& Lehnert 1997--hereafter WHL) 
and irregular galaxies (e.g. Martin 1997), is probably one of 
the ISM components that is most closely related to the feedback process.
The observed properties of the DIM are characterized by relatively 
strong low ionization forbidden lines compared to normal HII regions,
a low surface brightness, a rough spatial correlation with HII regions,
and a significant contribution (
$\sim$20\%--40\%) to the global H$\alpha$ luminosity. The importance
of this least studied ISM component
 is further emphasized by its likely ubiquity in galaxies
ranging from normal spirals to starbursts (Lehnert \& Heckman 1994;
Wang, Heckman \& Lehnert 1998). Studying the DIM, whose
properties are 
largely determined by the ionizing radiation and/or mechanical 
energy supplied by massive stars, 
will help us to better understand the interaction between
star forming regions and the ISM.
 
The existing observations raise many interesting questions about the physical
and dynamical state of the DIM that are still to be answered.
The energy required to power the DIM suggests that the gas either
soaks up nearly 100\% of the mechanical energy supplied by supernovae
and stellar winds, or the topology of the interstellar medium must
allow a substantial fraction of the ionizing radiation produced by
massive stars to escape HII regions and propagate throughout much of the
disk. The emission lines from the
DIM show relatively large linewidths compared to the
HII regions and in some cases, the
high ionization lines (e.g. [OIII]$\lambda$5007) have larger linewidths
than the low ionization lines (WHL).
This raises the intriguing possibility
that the DIM is not a uniform gaseous phase, and that different
heating mechanisms may play roles in different parts of the DIM.
WHL found there is a dominant, quiescent phase of the DIM which 
contributes more than 80\% of total H$\alpha$ emission of the
DIM and is likely photoionized by Lyc photons leaking from
HII regions, and a disturbed DIM component which is responsible
for the broadened high-ionization line emission and is probably
heated mechanically.
 
There are several pieces of evidence that favor 
photoionization of the quiescent
DIM by radiation leaking from HII regions (cf. WHL). Firstly, radiation
from massive stars is the only ionization source that can comfortably
provide the energy requirements of the DIM. Secondly, there is a morphological
correspondence between the DIM and giant HII regions 
(Hoopes et al. 1996; Ferguson et al. 1996; WHL). Thirdly, the 
transition from HII regions to the quiescent DIM is {\it continuous}
in physical and dynamical properties (e.g. WHL; Martin 1997). 
Fourthly, the relatively
narrow emission-lines of the quiescent DIM (WHL) argue against
mechanical heating as the dominant ionization source. Finally, 
photoionization models can readily explain the relative intensities
of the low-ionization lines in the quiescent DIM. These arguments
also rule out the possibility of other proposed energy sources for the
DIM, such as the decay of neutrino dark matter 
(Schiama 1990), which do not predict a relationship between the
DIM and HII regions. It is thus
essential to study the DIM in conjunction with the HII regions
that it surrounds.
 
An additional mystery about the relation between the DIM and HII
regions is the relatively constant fractional contribution of the
DIM to the global H$\alpha$ luminosity. Many studies show that roughly
20--40\% of the H$\alpha$ emission is from the diffuse gas, for different
types of galaxies (ranging from early-type spiral [Sb] to irregular) 
and different inclinations (Hunter \& Gallagher 1990; Kennicutt et al.
1995; Ferguson et al. 1996; Hoopes et al. 1996); this is inspite of the
fact that a variety
of methods were adopted to isolate the DIM. This roughly constant
DIM fraction suggests the DIM correlates with HII regions and therefore
the star formation process in a fundamental way. 
 
A crucial problem in the studies of the DIM is how to separate
the DIM from higher surface brightness gas (HII regions). Because
of their spatial correlation, the DIM seems more like an extension
of HII regions whose boundaries are not easy to define.
A simple approach is to set an absolute surface brightness limit
for the DIM. This method can not deal well with the case in which
mean surface brightness of the DIM varies across galactic disks (e.g.
Hoopes et al. 1996) or from galaxy to galaxy. Such variations will
occur due to differences in the heating rate of the DIM.
A criterion based on the {\it morphology} of the gas
is more important in this respect, because the diffuse nature of the
DIM is more fundamental than any seemingly arbitrary absolute surface
brightness. However, so far no such morphology-based methods have been 
developed to isolate the DIM.
 
In order to better understand the DIM, it is therefore very useful
to conduct a thorough investigation of its morphology,  
its relation with and differentiation from
 HII regions, and its
global importance. In this paper we present our analysis of H$\alpha$ images of
a sample of 7 of the nearest, largest, and brightest normal 
late-type galaxies to address these issues. The sample is selected
on the basis of Hubble type (Sb and later), proximity (closer than 
10 Mpc), large angular size ($>$ 10 arcmin), and relatively face-on
orientation (inclination $<$ 65$\arcdeg$), and includes M 51, M 81, M 101,
NGC 2403, NGC 4395, NGC 6946 and IC 342. Some general properties of
these galaxies are listed in Table 1. Our sensitive H$\alpha$
images of these galaxies can provide us with crucial information
about the structure and global significance 
of the DIM and, more fundamentally,
the common, basic properties shared by the DIM in star-forming galaxies.
 
\section{OBSERVATIONS AND DATA REDUCTION}

Most of the relevant observational information has been presented by WHL.
The only
change in our imaging data is that we now address the complete sample
including two new objects (NGC 6946 and IC 342) which were not discussed
by WHL. This imaging analysis is part of a larger project that investigates
the DIM in both the H$\alpha$ images and longslit spectra. Therefore in 
the following sections, we sometimes refer to the spectra, information
about which 
can be found in WHL, Wang, Heckman, \& Lehnert (1998), and Wang (1998).

Our imaging data were obtained during February 23-28, 1995 and November
1-2, 1994, with the 0.9 meter
telescope at KPNO. We used a Tek 2048$\times$2048 CCD with a 0.68\arcsec\  
pixel size, yielding a field of view of 23.2$\times$23.2 arcmin$^2$.
 The H$\alpha$
filter had a FWHM of 26$\AA$ and was centered on 6562$\AA$. 
Thus, relatively little flux from the [NII]$\lambda\lambda$6548,6584 
was included in these `on band' images. `Off-band' observations of the
continuum were made with a narrow band filter
having a FWHM of 85$\AA$ centered on 6658$\AA$. Several 1800(600) second
exposures were taken through the H$\alpha$ (continuum) filter
for each object.

  Data reduction was done using the IRAF package following standard procedures.
Images were bias-subtracted first and flat-fielded with a master flat combined
from a set of many dome flats. Sky-subtracted images were then geometrically
rectified to align H$\alpha$ on-band images with the off-band images.

The limit to the detection of diffuse, low-surface-brightness
H$\alpha$ emission in our data is set by the accuracy with which the 
contribution of continuum emission in the on-band image
can be subtracted (since the DIM H$\alpha$ emission has a small equivalent
width). In this regard, the images are inferior to long-slit spectra with
a spectral resolution matched to the intrinsic line-widths of the H$\alpha$
emission-lines (to maximize the contrast of the lines against the continuum).
We have therefore done this continuum subtraction in our images
 by scaling the off-band image such that difference image
agreed with the distribution
of the H$\alpha$ surface brightness in the long-slit spectra (as presented
by WHL and Wang 1998) in the
region of overlap. The agreement between the two sets of data 
has been ensured to be better than 10\%. 

To make this comparison,
we first sliced the continuum-subtracted H$\alpha$ line images 
in the regions covered by the
slit in the relevant spectral observations. We then reproduced
the spatial profile
of H$\alpha$ surface brightness variation in the same regions and compared it
to that shown by the spectra (see WHL and Wang 1998 for details about making 
similar profiles for the H$\alpha$ line in the spectra). By making small 
iterative 
adjustments to the scaling factor for the off-band image,
 we were able to get good matches of the profiles
between the
imaging and spectral data. This allowed us to obtain the final, most reliable
continuum-subtracted H$\alpha$ images.

Photometric H$\alpha$ imaging
observations of giant
HII regions(Kennicutt 1988) in these galaxies (except IC 342) 
were used to flux-calibrate
our images. We have also used our long-slit spectra to check for
consistency, and find that the agreement in H$\alpha$ flux in the regions
of overlap is
better than
10\% for NGC 2403, NGC 6946, NGC 4395 and M 101 and 30\% for M 51. 
For IC 342, we used the flux scale determined from our long-slit spectra
of H$\alpha$,
which were taken under photometric conditions, to flux-calibrate
the image. The result was then checked against the flux calibration
provided by narrow-band images of
standard stars. We find a consistency better
than 30\%. This indicates the flux calibration for the IC 342 image
is also reliable. 
 
Internal extinction in the galaxies has not been corrected for. 
For most of the galaxies, foreground Galactic extinction is not 
important. From the HI survey by Stark et al. (1992), we
derived the Galactic extinction in magnitudes at 
H$\alpha$ $A_{H\alpha}$ (Table 1).
The Galactic HI column densities are substantial
only in the directions toward NGC 6946 
and IC 342: they imply $A_{H\alpha}$ to be 1.63 for IC 342 and
1.43 for NGC 6946. These values are consistent with the previous measurements
of $E_{B-V}$ within the uncertainty: $A_{H\alpha}$ was 
found to be 1.8  for IC 342
(McCall 1989)
and 1.0 for NGC 6946 (Burstein \& Heiles 1984) if $A_{H\alpha} = 2.5 E_{B-V}$.
Using our values, the fluxes at H$\alpha$ for these two galaxies
would be raised by a factor of $\sim$4 after foreground extinction correction.

\section{RESULTS}

The H$\alpha$ images of our sample galaxies are shown in Figure 1. 
The images have been smoothed with a box having a size of
2$\arcsec\times$2$\arcsec$.
A faint, diffuse gas component can be seen throughout the disks of 
these galaxies. The detection of extremely faint
gas is affected by the uncertainty of continuum subtraction of
the images. But as discussed by WHL and Wang (1998), the wide spatial
distribution
of the DIM has also been confirmed by our spectra, which have
a higher sensitivity.
The diffuse gas is preferentially located along the edges
of spiral arms and around isolated HII regions, as is generally the case
for the DIM (e.g. Ferguson et al. 1996; Hoopes et al. 1996). 
For M 81, the
diffuse gas component within a radius of 2--3\arcmin\ from the nucleus
seems to not be associated with any prominent star-forming regions and
is probably different in nature from the DIM seen elsewhere. The
LINER-type nucleus in this galaxy or a population of hot post-main-sequence
low-mass stars may play a role in energizing the gas (Devereux, Jacoby, \&
Ciardullo 1995, 1996).

  \subsection{Growth Curves}

  A simple approach to study the DIM contribution to the total H$\alpha$
flux is to plot the integrated flux against a surface brightness cutoff below which the
flux is integrated (a `growth curve'). Instead of clipping out discrete HII 
regions before smoothing the image, as done by WHL in order to avoid
contamination of bright HII regions to the DIM, here we use a more
straightforward method of simply smoothing the raw image with a 
2$\arcsec\times$2$\arcsec$ box and constructing a growth curve. 
This method results in slight differences from the results in WHL,
but the robust pattern is still the same.

Because the typical surface brightness of the DIM ($\sim$10 pc cm$^{-6}$)
\footnote{Throughout the paper we convert surface brightness to emission
  measure assuming an electron temperature of 10$^4$ K.
  5$\times$10$^{-17}$ erg s$^{-1}$ cm$^{-2}$ arcsec$^{-2}$ 
  corresponds to 25 pc cm$^{-6}$. }
is comparable to the noise of the background of the resulting smoothed 
images, we 
determined the flux from these images by integrating
from $-3\sigma$ below the mean background of the images up to a given 
surface brightness level. We repeated this
process over a range of surface brightness, starting at a level of
3$\sigma$
above the background of the smoothed images.
This minimum surface brightness level corresponds to an
emission-measure (foreground extinction corrected) of
55 pc cm$^{-6}$ for IC 342 and
NGC 6946, 30 pc cm$^{-6}$ for
M 51 and M 101, 10 pc cm$^{-6}$ for M 81 and NGC 2403, and 20 pc cm$^{-6}$
for NGC 4395. The depths of the images do not significantly affect the
overall result.
Figure 2 thus shows the fraction of the total H$\alpha$ flux
contributed
by gas below a given surface brightness (emission measure). The
data for NGC 6946 and 
IC 342 in Figure 2 have been corrected for foreground Galactic
extinction (see above). 

There are four effects that should be noted when
interpreting Figure 2. First, the light from
HII regions, scattered by the optics in the telescope and camera 
and by the dust in the galaxies, has not been corrected for.
Hoopes et al. (1996)
showed that this correction is minor for their sample galaxies. Furthermore,
the emission-line ratios of the DIM are significantly 
different from that of typical HII regions in these galaxies (WHL; Wang 1998).
No scattering
processes 
can account for this difference. Thus, scattered light can not dominate the
DIM flux and we
do not consider this effect here. Second, a more reliable determination 
of the DIM contribution
should include a correction of the missed diffuse gas flux projected on top of
regions of higher surface-brightness (HII regions). A simplified method of
 making this correction
(by assuming the diffuse gas in the clipped-out regions
has the mean surface brightness level measured elsewhere, e.g.
 Ferguson et al. 1996;
Hoopes et al. 1996)
may increase the DIM fraction significantly, especially for low cut-off levels.
At 50 pc cm$^{-6}$, the correction factor may be up to 2 (Hoopes et al. 1996).
Third, for the purpose of our study, we did not indicate 
the uncertainty in continuum subtraction in Figure 2.  Based on the comparison
of our spectra and images, this uncertainty is minor at the relevant levels
of emission measure in most cases,
and will not alter the essential behavior of the growth
curve. M 101 is the only object whose growth curve is relatively
uncertain at low surface brightness ($<$ 100 pc cm$^{-6}$)
as caused by continuum subtraction. This is mainly due to two 
reasons. The image of M 101 is almost filled by the galaxy itself, making
it hard to determine a proper background level to subtract. 
In addition, the DIM in M 101 is less prominent compared to the DIMs
in the other galaxies so the S/N ratio 
for the measurement of the M 101 DIM is relatively poor.
Fourth, no internal extinction correction has been applied. In a
situation in which HII regions suffer more average extinction than the DIM,
the intrinsic DIM fraction should drop accordingly. Because of the small
EQWs of the H$\alpha$ and H$\beta$ emission lines in the DIM, we were not
able to
accurately determine the extinction for the DIM using the Balmer decrement
from our spectra (see WHL and Wang 1998 for the spectra).

  The growth curves shown in Figure 2
 vary among our sample galaxies. In
M 51, M 101, NGC 2403, NGC 6946 and IC 342 the flux contribution
of the DIM is small relative to that in
NGC 4395 and M 81. 
We verified that the high DIM fraction in M 81 is not due to 
the prominent diffuse gas component in the inner disk or bulge. 
To do this test, we masked out a circular region within a radius 
of 2.5\arcmin\ 
from the nucleus and constructed the growth curve again. The result showed
little difference from that in Figure 2. Thus, the high fraction of the DIM 
contribution is purely due to the fact that the DIM component in
M 81 is intrinsically prominent throughout the galaxy. 

    \subsection{The Surface-Brightness Distribution Function}

Although we (and others) have defined the DIM in terms of a surface-brightness
criterion, the D in DIM stands for `Diffuse'. Thus, the information
on the areal coverage
of low H$\alpha$ surface brightness gas is as important as the flux
contribution 
from the same gas.
We make the observations that both the areal and flux contributions
from the gas at certain H$\alpha$
surface brightness levels are determined uniquely by a distribution
function $A(\lg\Sigma)$ that describes the number of pixels occupied by gas
with H$\alpha$ surface brightness within an interval of $d\lg\Sigma$ centered at $\lg\Sigma$.
It is therefore worthwhile to construct this function to quantify
the `diffuseness' of the DIM.
 
Since we can only measure the area occupied by gas with a given surface 
brightness down to the detection limit in an image (as characterized by
+3$\sigma$ above the background) and our images have slightly
different sensitivities, we chose a common value for this minimum $\Sigma$,
which is no
less than +3$\sigma$ above the background in the relevant images.
We have smoothed the images 
with a 4$\arcsec\times$4$\arcsec$ box before
measuring the area at different surface brightness levels. 
This smoothing with a larger box allows us to measure down to a minimum
surface brightness limit that is half as high as the limits given in
\S3.1 for a 
2$\arcsec\times$2$\arcsec$ box smoothing. 
We then chose a common depth corresponding to a surface brightness of
3$\times$10$^{-17}$ ergs s$^{-1}$ cm$^{-2}$ arcsec$^{-2}$
(an emission measure of 15 pc cm$^{-6}$)
for M 51, M 81, M 101, NGC 2403 and NGC 4395, and
6$\times$10$^{-17}$ ergs s$^{-1}$ cm$^{-2}$ arcsec$^{-2}$
(an emission measure of 30 pc cm$^{-6}$)
for NGC 6946 and IC 342 after foreground extinction correction.
The images of NGC 6946 and IC 342 have poorer sensitivities due to
foreground extinction, so we set a higher surface brightness limit for them.
This procedure removes the arbitrary
factor of the relative depth of the images. 

Following these ideas, we have constructed the distribution
function $A(\lg \Sigma)$ in Figure 3. 
We also measured the area within the galaxy disk lying above
the minimum limiting surface brightness (see above) and 
the mean surface brightness 
$\bar{\Sigma}$ of H$\alpha$ within this area.
The area actually used constitutes about 10\%--40\% of the total disk area
lying within the galaxy's
25 B mag arcsec$^{-2}$ isophote, 
and is illustrated in Figure 4 for
M 51 and M 101. These measured area and $\bar{\Sigma}$ (as listed in
Table 2) 
were used respectively to scale $A$ and $\Sigma$ when
plotting the distribution function. The purpose of such a scaling is to
eliminate any galaxy-dependent factors in the relation between A and $\Sigma$,
and thus allow meaningful comparisons between galaxies with very different
surface areas or mean surface-brightnesses. Note that,
unlike what we did for the growth curves, we now exclude
the data with surface brightness 
between $-3\sigma$ of background and our adopted minimum isophotal limit.
The reason is that the morphological
properties of the data near the noise level of the images
is not well defined.  
For instance, in M 101 the area 
above the isophotal limit including both the diffuse gas and
HII regions constitutes $\sim$10\% of 
 the total disk area in contrast to $\sim$35\% for M 51 
(Table 2 and Figure 4). 
This difference in the projected fractional area  
of bright emission-line gas is likely 
related to arbitrary factors such as different inclination angles of 
the two galaxies (compare 64$\arcdeg$
for M 51 and 17$\arcdeg$ for M 101). When measuring 
the mean surface brightness and covering area of the DIM, 
it is then meaningful to only use
the area with a surface brightness above the detection limit. 

Figure 3 shows the curves lie along almost the 
same locus, especially in the regime of $\Sigma/\bar{\Sigma} \gtrsim$ 1, 
regardless of objects. We infer that the curves at high surface 
brightness can be
approximately characterized by an universal function 
\begin{equation}
  \frac{A(\lg \Sigma)}{A_{tot}} = F(\frac{\Sigma}{\bar{\Sigma}})
\end{equation}
At lower surface brightness ($\Sigma/\bar{\Sigma} \lesssim$ 1), the curves 
start to diverge, i.e.
the distribution functions for M 51, M 101, NGC 2403, NGC 6946 and IC 342 have
shallower faint-end slopes 
compared to those of M 81 and NGC 4395. 

The different behavior around the turning point $\Sigma/\bar{\Sigma} \sim 1$
may not be coincidental. 
We note that $\Sigma = \bar{\Sigma}$ is roughly the dividing point
between the DIM and HII regions for two reasons. Firstly, the mean surface 
brightness we measured
(Table 2)
roughly matches with the limiting surface brightness that we have previously
used to
separate the DIM from HII regions
in the spectra ($\lesssim$25 pc cm$^{-6}$ for the DIM and
$\gtrsim$100 pc cm$^{-6}$ for bright HII regions; see WHL).
We have shown in our analysis in WHL that these criteria are
effective in discriminating between the 
spectroscopic properties of the DIM and HII regions.
Secondly, the universal scaling relation (eq. [1]) for the high 
surface brightness portion of the distribution function
can be derived from the luminosity function and 
radial brightness profiles of the HII regions (to be discussed in \S4.1), 
and it is therefore solely determined by the physics of high-mass star
formation in galaxies. On the other hand, the varying slope of 
the curve at $\Sigma/\bar{\Sigma} \lesssim$ 1 must reflect some intrinsic
difference in the DIM properties in galaxies.
Therefore, the choice of the mean surface brightness to separate the DIM
from the HII regions is
sound  physically. The advantage of separating the DIM from HII regions
based on $\bar{\Sigma}$ is that it takes into account the information
about the {\it relative}
 brightness of the emission-line gas in a galaxy (i.e. how
bright the gas is typically) instead of only relying on some arbitrary
absolute surface brightness criterion.

Although relative surface brightness may be an effective 
criterion to separate the DIM
from HII regions, we point out that 
the critical value for $\Sigma/\bar{\Sigma}$ does not have to
be exactly unity. A general dividing point should be defined as
$\Sigma/\bar{\Sigma} = const$ where the constant 
depends on the value of the minimum surface brightness chosen to delimit
the area of the galaxy within which $\bar{\Sigma}$ is determined.
That is, more sensitive
imaging would allow the value for $\bar{\Sigma}$ to be determined
over a greater fraction of the total disk area by making measurements
possible in regions with smaller surface brightnesses.

However, we should point out that Figure 3 will still be rather robust to
such effects (i.e. the distribution function will be rather insensitive to
the value of the common isophotal limit we have adopted
for our images).
Although $A(\lg \Sigma)$  does not  rely on the
lowest surface brightness we can measure, 
the mean surface brightness and total area that we use to 
scale the plot do 
depend on it. We found a drop in the lower bound of surface 
brightness increases 
the total area
significantly but has little impact on the total flux, which is dominated
by high surface brightness gas (i.e. HII regions). As a 
result the area will increase while
$\bar{\Sigma}$ will decrease by roughly the same factor.
A shift of the curve in Figure 3
 due to change in $\bar{\Sigma}$ would be approximately compensated by a
shift due to total area change.
We conclude that the scalability of the distribution function
would remain roughly unaffected if the common isophotal limit
is varied. This is supported by the similarity of the curves
between the low-Galactic-latitude galaxies NGC 6946 and IC 342 (having a higher limiting isophotal level 
in the images) and
the rest of our sample galaxies (Figure 3).

\subsection{Integrated Flux and Area Contribution of the DIM}

It is worthwhile to examine the growth pattern of not just the H$\alpha$ flux
but
also the surface area occupied by the DIM. Thus, we plot
 the fractional integrated flux against fractional integrated area in
Figure 5 using exactly the same surface brightness thresholds
and method of measurement as discussed in the previous section. 
We emphasize that the result in
Figure 5 should be 
completely determined by the surface-brightness
distribution function (Figure 3).
Because the relative surface brightness is an important
parameter to discriminate the DIM from HII regions (see above),
we mark the curves with the ratio of $\Sigma/\bar{\Sigma}$ 
(instead of absolute surface brightness), where
the symbols represent successive
increases in surface brightness by a factor of 1.25
 from ${\Sigma}/{\bar{\Sigma}} = 0.4$
 (lower left corner) to
${\Sigma}/{\bar{\Sigma}} = 3$ (upper-right corner). 

Figure 5 shows that the gas with surface brightness from 
the limiting
surface brightness to a given cutoff level of 
${\Sigma}/{\bar{\Sigma}}$ contributes to the global H$\alpha$ flux and 
covering area in a similar way in all the galaxies, especially at high
surface brightness end. There is a difference in the contribution from
the fainter gas (the DIM), e.g. at a fixed area fraction, the DIM
in M 81 and NGC 4395 contributes more to global flux than the
rest of the galaxies. 

Figure 2, Figure 3, and Figure 5 are
complementary ways of quantifying the contribution of the DIM, and all
yield a consistent picture. At high levels of surface-brightness
(the HII regions), all the galaxies are very similar. Differences become
apparent at low surface-brightnesses, where the
DIM is
more significant in M 81 and NGC 4395, and less significant
in M 51, M 101, NGC 2403, NGC 6946 and IC 342.
 
Figure 5 graphically illustrates why the measured fractional flux 
contribution from the DIM is roughly constant ($\sim$20\%--40\%), 
regardless of 
Hubble type and other factors (e.g. Ferguson et al. 1996; 
Hoopes et al. 1996). As ${\Sigma}/{\bar{\Sigma}} \sim 1$ 
is probably the point where
HII regions and the DIM start to differentiate, we can use this criterion
to measure the total DIM contribution to the global H$\alpha$ flux. We
then find from Figure 5 that
this contribution converges uniformly at $\sim$20\%--40\% regardless
of galaxies.  These values have not been corrected for any
DIM emission on top of HII regions that might have been
missed in the measurement. The fraction is 
consistent with the measurements from previous studies.
Our conclusion is obtained by isolating  the DIM based on
the distribution function behavior (and therefore fundamentally on
H$\alpha$ morphology)
 instead of picking an arbitrary absolute surface brightness criterion
as others adopted previously. So we have confirmed that the contribution
from the DIM to the global flux is indeed {\it roughly} constant
and there exists a physically-meaningful explanation for this: 
although the slopes of the distribution function of the DIM for different
galaxies may vary, they are still similar enough that the flux 
and area contribution of the DIM converges to a roughly common value.
It seems that the morphological information has been
more or less taken into account when choosing
the absolute surface brightness criterion in previous studies.  
As a result the DIM was
appropriately isolated and the measurements of its flux contribution 
were roughly correct.

\section{DISCUSSION}

\subsection{The Origin of the Scalable Distribution Function}

It is important to understand the underlying physics of the 
universal distribution function for HII regions 
($\Sigma/\bar{\Sigma} >$ 1) because it should be physically related to the DIM
properties that we are exploring. Statistically, we need to 
consider the luminosity function and size function of HII
regions, as these two functions determine the bright end of the distribution
function.
The common properties of HII regions are the reason that different
galaxies have approximately the same curve in Figure 3 
at high levels of surface brightness, as we now show.

  We model HII regions as isolated (non-overlapping) sources, 
each of which has an exponential H$\alpha$ surface brightness profile 
(e.g. Rozas,
Casta$\tilde{\rm n}$eda and Beckman 1998; Kennicutt 1984)
        \begin{equation}
        \Sigma(\theta) = \Sigma_0 \ e^{-\frac{\theta}{\theta_0}}
        \end{equation}
where $\Sigma(\theta)$ is the surface brightness at a given
angular distance $\theta$, $\Sigma_0$ is the peak surface brightness
at center, and $\theta_0$ characterizes the angular extent of the
HII region. 
$\Sigma_0$ is then related to luminosity L and $\theta_0$ as
        \begin{equation}
        \Sigma_0 \propto \frac{L}{\theta_0^2}
        \end{equation}

  There are many studies regarding the luminosity function
of HII regions. A simple power-law usually holds well for many
galaxies, while a broken power-law is better for some others
(e.g. Kennicutt, Edgar and Hodge 1989; Oey \& Clarke 1998). Here we adopt a
power-law
with a constant power index as 
        \begin{equation}
        N(L) dL \propto L^{-\beta} dL
        \end{equation}
Although the size distribution for HII regions has been known to be
exponential ($N(\theta_0) \propto \exp(-\theta_0/\theta_c)$, e.g. van den Bergh 1981, where $\theta_c$ can represent the characteristic angular size
spread of the HII regions),
we found the detailed functional dependence of $N(\theta_0)$
on angular size $\theta_0$ does not affect
our following arguments. So here we only assume there exists a universal
$\theta_0$ distribution of the form
        \begin{equation}
        N(\theta_0) d\theta_0 = f(\theta_0) d\theta_0
        \end{equation}
where $f(\theta_0)$ is some function of $\theta_0$. In addition,
we assume that L, $\theta_0$ and $\Sigma_0$ distribute independently.
For example, for HII regions with 
$\theta_0$ within any given interval the luminosity
distribution still follows eq.[4].
Based on eq.[3] and eq.[4], we find the $\Sigma_0$ distribution 
obeys a power-law similar to the luminosity function
        \begin{equation}
        N(\Sigma_0) \, d\Sigma_0 \propto \Sigma_0^{-\beta} \, d\Sigma_0
        \end{equation}

  Each source has a contribution to covering area (absolute value in steradian)
at a given surface brightness interval of 
        \begin{equation}
        A(\lg \Sigma) \equiv \frac{dA}{d \lg\Sigma} = 2 \pi \theta_0^2  \,
          \lg \frac{\Sigma_0}{\Sigma}
        \end{equation}
The total amount of $A(\lg \Sigma)$ contributed by all HII regions 
in a galaxy  is a sum of the contribution by HII regions with a given
angular size $\theta_0$ and a given peak surface brightness $\Sigma_0$,
i.e. it is simply $A(\lg \Sigma)$ 
integrated over the ranges of 
$\Sigma_0$ and $\theta_0$.
The result of integrating over $\Sigma_0$ depends on the surface
brightness $\Sigma$ that we are considering, but the integral of 
$\theta_0$ does not. 
Since we know the number distributions of
HII regions with angular size (eq.[5]) and peak surface brightness (eq.[6]),
it can be shown that the total area for a galaxy
at a given surface brightness is 
        \begin{equation}
        A_{tot}(\lg \Sigma) = \zeta \int_{\Sigma}^{\Sigma_{max}} d\Sigma_0 \,
          \Sigma_0^{-\beta} \, \lg \frac{\Sigma_0}{\Sigma}
        \end{equation}
where $\Sigma_{max}$ is the highest peak surface brightness among the
 HII regions
(corresponding to the highest surface brightness 
we can measure in the image if the sources are not overlapping), which can
be assumed to be the same for all galaxies as a good approximation,
and $\zeta$ is a factor containing the result after integrating over
$\theta_0$ and other quantities including
distance from the Milky Way, total number of HII regions and
characteristic angular size spread $\theta_c$. 
Eq.[8] implies that as long as $\Sigma$
is not comparable to $\Sigma_{max}$ (i.e. $\Sigma \ll \Sigma_{max}$),
$A_{tot}(\lg \Sigma) \propto \Sigma^{1-\beta}$. Therefore, the slope
of the distribution function tracks that of the luminosity function. For
$\beta$ = 2$\pm$0.5 as commonly observed, the range of the
distribution function slope  encloses the observed values
in Figure 3. To be more specific, 
the dependence of $A_{tot}(\lg \Sigma)$ on $\Sigma$
(eq.[8]) is plotted on Figure 3 (solid line, $\zeta$ is arbitrarily
chosen) for which we adopt $\beta$ = 2 and $\Sigma_{max}/\bar{\Sigma}$ = 100,
appropriate for our images.

  The model prediction (eq.[8]) fits our data in
Figure 3 rather well at high surface-brightness. This implies the
universal H$\alpha$ distribution
function for the bright gas is probably just a manifestation of more
fundamental HII regions properties such as luminosity function and
radial exponential brightness profile (eq.[2] and [4]), which are
shared by late-type galaxies and which reflect the underlying star-formation 
process (e.g. Oey \& Clarke 1998). Indeed, eq.[8] dictates that the 
galaxy-specific $\zeta$ parameter is cancelled out in the relation
between $A_{tot}(\lg \Sigma) / A_{tot}$ and $\Sigma / \bar{\Sigma}$,
which only depends on the minimal and maximum surface brightness 
that can be measured. In our case we can regard these limiting surface
brightnesses as the same for all sample galaxies. Therefore the
$A_{tot}(\lg \Sigma) / A_{tot}$ vs. $\Sigma / \bar{\Sigma}$ relation
should be universal among galaxies just as observed.

The data at very high surface brightness 
($\Sigma / \bar{\Sigma} \gtrsim$
10--20, Figure 3) show a relatively larger scatter among 
galaxies. There might be two major reasons for this. Firstly, the 
luminosity function can be relatively steep for some galaxies  at 
high L due to the broken power-law (e.g. Kennicutt et al. 1989). 
It is possible that some of our sample galaxies 
(e.g. M 51, Kennicutt et al. 1989) may belong to this 
category. 
This may result in a steeper slope of the 
distribution function at high surface brightness. 
Secondly, as the number of pixels at very high 
surface brightness is small, statistics
may be affected by bad pixels in the images which may be caused by,
e.g., residuals of bright stars after continuum subtraction.

\subsection{Similarities and Differences in Global DIM Properties Among
Galaxies}

The universal distribution function of HII regions determines that the
emission line nebulae are scalable in surface brightness for $\Sigma >
\bar{\Sigma}$. 
The DIM can then be separated
from HII regions according to relative surface brightness 
($\Sigma < const \times \bar{\Sigma}$, 
where {\it const} $\sim$ 1 for the limiting surface brightness 
reached by our images).
This criterion is morphologically-based and more meaningful than an absolute
surface brightness limit. For galaxies with relatively high
average surface brightness, such as starbursts,
we expect the DIM surface brightness criterion would be higher too. 
From comparison of
emission-line properties between the normal galaxies discussed in this paper
and the Lehnert \& Heckman (1995) sample of starbursts,
Wang, Heckman and Lehnert (1998)
indeed found a DIM component in starbursts that has physical
properties similar to those in our normal galaxies, except that 
the starburst DIM
has relatively large absolute surface brightness
compared to the DIM 
in normal star-forming galaxies. In fact, these properties depend only on
a relative surface-brightness parameter defined as the surface brightness
scaled by the mean surface brightness $\Sigma_e$ within the
galaxy's H$\alpha$ half light
radius $r_e$ (as given in Table 2). Since $\Sigma_e$ is directly proportional
to the mean rate of star formation per unit area in the disk, this result
is not surprising if we consider the very high star formation rate for
the starbursts. This
supports our argument here that the {\it relative} 
surface brightness ($\Sigma / \bar{\Sigma}$)
is the fundamental property by which to define the DIM.
On the other hand, it is worth noting that the DIM in our normal
galaxy sample as defined by the mean surface brightness is still 
a factor of 10$^1$--10$^2$ brighter than the Reynolds layer
in our Galaxy, which has a typical emission-measure of only
$\sim$1--5 pc cm$^{-6}$ (Reynolds 1990).

Figure 3 shows that  the
transition from the high surface brightness gas 
(HII regions) to the low surface brightness gas (the DIM)
is smooth. Therefore there must be a strong relationship 
between these two otherwise
an abrupt  change in $A(\lg \Sigma)$ below $\Sigma/\bar{\Sigma}\sim$1 
would be possible. This is consistent with the picture of a smooth
transition in physical properties from HII regions to the DIM (WHL)
and thus
provides additional support for the idea
that a unified approach to the study of both the DIM and HII regions
is necessary to better understand the DIM.

In the low surface brightness range 
($\Sigma/\bar{\Sigma} < 1$), the DIM in some galaxies such as
M~101 covers less fractional area than other galaxies at a given 
$\Sigma/\bar{\Sigma}$.
Although we are not able to 
measure surface brightness far below 
$\bar{\Sigma}$ to fully assess the variation of the 
distribution function at low surface brightness end, 
we find that the data in Figure 3
can all be fitted well with a double power law with a break around 
$\Sigma/\bar{\Sigma} \sim 1$. While the slope of the distribution function
for HII regions ($\Sigma/\bar{\Sigma} >$ 1) is approximately the same,
it is not the case for the DIM at low surface brightness. 
Therefore we suggest using the 
different power-law indices measured for the region of 
$\Sigma/\bar{\Sigma} <$ 1
 to indicate the prominence of 
the DIM in each galaxy. By this criterion, the DIM is most pronounced in
M 81 and NGC 4395 (see Table 2). 

It is not clear yet what causes the difference in the
distribution function of the DIM. 
It is also possible that as 
the sensitivity of the images improves, we would find more significant
variation 
in the DIM distribution function at still fainter levels, and as a result more
substantial
differences in the global properties of the DIM among galaxies.

  WHL have tried to
characterize the relative importance of the DIM in each galaxy using a
measurement that involves only the structure or morphology of the H$\alpha$
images. We have characterized the `DIMness'
of the galaxy by taking the ratio of the mean and the r.m.s. of the H$\alpha$
surface-brightness (computed for surface brightness 
$\gtrsim$ $-3\sigma$ of background) 
within the surface area of the galaxy delimited by
the 25 B magnitude per square arcsec isophote. This ratio may be affected by
the arbitrary fractional area of  regions
with surface brightness below the sensitivity level as discussed in \S3.2.
However, we should expect this
mean and r.m.s. ratio to be fairly constant 
if we only compute it
for regions with surface 
brightness above our common limiting isophotal level instead,
as the `DIMness' parameter then solely depends on the distribution 
function in Figure 3. 
Indeed, we found that this is the case.
Therefore, the ratio of the mean and r.m.s. 
of the H$\alpha$ surface brightness may not
be a good indicator of the DIM prominence in 
galaxies.

\section{CONCLUSIONS}
 
We have reported on the results from a program to study
the structure and morphology of a sample of 
nearby bright normal late-type galaxies in order to understand the global
properties of the DIM. 
For each of the seven galaxies (NGC 2403, M 81, NGC 4395, M
51, M 101, NGC 6946 and IC 342) 
we have analyzed deep narrow-band H$\alpha$ images
covering essentially the entire star-forming disk (a field diameter
of 23.2 arcmin, or 18 to 53 kpc). These images reach limiting
H$\alpha$ surface-brightnesses of about 2--10 $\times 10^{-17}$ erg
cm$^{-2}$ s$^{-1}$ arcsec$^{-2}$ (corresponding to an emission
measure of about 10--50 cm$^{-6}$ pc).

While the DIM contribution to the global
H$\alpha$ luminosity seems different when the DIM is defined by a
given absolute H$\alpha$ surface
brightness, we found that the DIM properties
can be reasonably unified as a function of relative
surface 
brightness ($\Sigma / \bar{\Sigma}$, where $\bar{\Sigma}$ is the mean
surface-brightness within regions lying above a fixed
very faint isophotal level).

The DIM structural properties
are described by a fundamental dimensionless
`distribution function' that measures
the {\it relative} area within the galaxy covered by gas at a given 
{\it relative} surface
brightness. Indeed, we found this
function is very similar for all sample
galaxies, especially at high-surface-brightnesses
($\Sigma / \bar{\Sigma} > 1$). We show that this behavior
at the high-end is determined only by common HII region properties
(the
H$\alpha$ luminosity function and the radial brightness
profiles of HII regions).

At lower surface brightness ($\Sigma / \bar{\Sigma} < 1$),
the distribution function becomes more diverse, indicating a variation in
the DIM prominence among galaxies. In some galaxies (M 51, M 101, NGC 2403,
NGC 6946 and IC 342) the distribution function shows a pronounced flattening
in power-law slope at the faint-end, indicating a less-conspicuous DIM.
In the other galaxies (M81 and NGC 4395), the faint-end 
slope
is steeper (similar to the bright-end slope), and so the DIM is more
pronounced. This method and
 two types of growth curves we made using different methods yield
a consistent picture about the global importance of the DIM in our sample
galaxies, and they are all valuable in providing
complementary
information about the DIM.

These results
imply that the 
DIM should be defined by using 
the relative 
H$\alpha$ surface brightness
criterion $\Sigma / \bar{\Sigma} \sim$ 1 as the boundary between the DIM
and HII regions.
This criterion is free of factors
such as the varying average H$\alpha$ surface brightness 
(mean star-formation rate per unit area) from one galaxy to another,
and is more
physically sound than an absolute surface brightness limit.

The distribution function shows the continuous transition from
HII regions to the DIM, suggesting a tight coupling between
these two gaseous phases. This is consistent with the idea 
that the majority of the DIM is photoionized by Lyman continuum photons
leaking from HII regions. We found that while some variation
in the global importance
of the DIM exists among galaxies,
a structural (morphological) similarity of the DIMs in our
sample galaxies still exists because of the roughly similar distribution
functions. This similarity dictates that the
fractional
 contribution of the DIM to the global H$\alpha$ luminosity in a galaxy
is fairly constant (within a factor of $\sim$ 2)
when the DIM is isolated based on its relative surface brightness.

Our related paper (Wang, Heckman, \& Lehnert 1998) extends these ideas
about the DIM, and shows that normal star-forming
galaxies and starbursts can be unified if the DIM is defined
in terms of relative surface brightness. Taken together, our work
suggests that the DIM is indeed a fundamental component of all
star-forming galaxies, that it is ionized primarily by photons produced
by high-mass stars, and that the global structure of the diffuse
ISM in star-forming galaxies regulates and/or
is regulated by the mean rate of star-formation
per unit area in the galaxy.

\acknowledgments
The work of M.D.L. at Leiden is supported by a program funded by the Dutch
Organization for Research (NWO) and the Dutch Minister of Education and
at IGPP/LLNL (where portions of this work were completed) under the
auspices of the US Department of Energy under contract W-7405-ENG-48.

\clearpage


\begin{planotable}{lcccccc}
\scriptsize
\tablewidth{38pc}
\tablecaption{Galaxy Parameters} 
\tablehead{
\colhead{Object }      & \colhead{Hubble Type\tablenotemark{a}} &
\colhead{$i$\tablenotemark{b}} & \colhead{D(Mpc)\tablenotemark{c}}    &
\colhead{$v$(km/s)\tablenotemark{d}}   &  \colhead{A$_{{\tt H}\alpha}$\tablenotemark{e} } &
\colhead{M$_{B,T}$\tablenotemark{f}  }
}

\startdata

M 51     &  SbcI-II  &       64\arcdeg   &     8.4     &   464   &  0.07  &
-21.0 \nl
M 81     &  SbI-II     &       60\arcdeg   &    3.6     &   -36   &  0.23  &
-20.8  \nl
M 101    &  ScdI     &       17\arcdeg   &     7.4     &   251  &  0.07  &
-21.5 \nl
NGC 2403 &  ScIII     &       62\arcdeg   &     3.2     &   131  &  0.25   &
-19.2 \nl
NGC 4395 &  SdIII-IV   &    38\arcdeg   &     2.6     & 317   &  0.07  & 
-16.7 \nl
NGC 6946 &  ScII     &   42\arcdeg   &   5.9   &    48   &   1.43  &  -21.5 \nl
IC 342   &  Scd      &   20\arcdeg   &   2.1   &    32   &   1.63 &  -20.2 \nl

\tablenotetext{a}{ From the Revised Shapley-Ames Catalog(Sandage \& Tammann 1987)
        except that IC 342 type is from Tully(1988).}
\tablenotetext{b}{ Inclination data are from Tully(1988) except the 
        inclination for M 101
        is from Zaritsky et al. (1990). }
\tablenotetext{c}{ Distances are from Feldmeier et al. (1997) (M 51),
        Freedman et al. (1994) (M 81), Kelson et al. (1996) (M 101),
        Karachentsev \& Makarov (1996) (NGC 2403), Rowan-Robinson (1985)
        (NGC 4395), McCall (1982) (NGC 6946) and 
        Karachentsev \& Tikhonov (1993)(IC 342). }
\tablenotetext{d}{ Weighted mean observed heliocentric radial velocity from
        Revised Shapley-Ames Catalog(Sandage \& Tammann 1987)
        except that $v$ for IC 342 is from Tully(1988).}
\tablenotetext{e}{ Galactic extinction in magnitudes at H$\alpha$,
        estimated based on the foreground Galactic HI column
        density(Stark et al. 1992) in the direction of the objects. 
        The conversion is done by assuming
        N$_{HI}$/E$_{B-V}$ = 5$\times$10$^{21}$ cm$^{-2}$ mag$^{-1}$ and 
        A$_{{\tt H}\alpha}$ = 2.5 E$_{B-V}$. }
\tablenotetext{f} { Total absolute magnitude in the B band based on the apparent
        magnitude from the
        Revised Shapley-Ames Catalog(Sandage \& Tammann 1987)
        (except that IC 342 M$_{B,T}$ is from Tully(1988)), 
        corrected to our adopted distance. These values have
        been corrected only for foreground Galactic extinction. }

\enddata

\end{planotable}



\begin{deluxetable}{lcccccccl}
\scriptsize
\tablecaption{DIM H$\alpha$ Imaging Results}
\tablehead{
\colhead{Object}   &
\colhead{H$\alpha$ Flux \tablenotemark{a}}      & 
\colhead{L$_{{\tt H}\alpha}$ \tablenotemark{b}} &
\colhead{$\bar{\Sigma}_{{\tt H}\alpha}$\tablenotemark{c}} &
\colhead{$\Sigma_e$ \tablenotemark{d}} &
\colhead{r$_e$ \tablenotemark{e}} &
\colhead{$\frac{A_{used}}{A_{true}}$ \tablenotemark{f}} &
\colhead{a$_{25}\times$b$_{25}$ \tablenotemark{g}}   &
\colhead{$n$ \tablenotemark{h}}        \\
\colhead{} & \colhead{(erg s$^{-1}$ cm$^{-2}$)} & \colhead{(L$\sun$)} & 
\colhead{(pc cm$^{-6}$)} &
\colhead{(pc cm$^{-6}$)} &
\colhead{(kpc)} &
\colhead{} &
\colhead{(kpc)} &
\colhead{} }
\startdata
M 51  &  2.0$\times$10$^{-11}$  &  4.4$\times$10$^{7}$  & 130  &  
     120  &  4.2  &  0.35  &  27.4$\times$16.9  &  0.42   \nl
M 81  &  3.7$\times$10$^{-11}$  &  1.5$\times$10$^{7}$  & 75  &  
     30   &  4.2  &  0.20  &  28.2$\times$14.8  &  1.1    \nl
M 101 &  3.7$\times$10$^{-11}$  &  6.3$\times$10$^{7}$  & 105  &  
     35   &  10.2 &  0.13  &  62.0$\times$57.9  &  0.43   \nl
NGC 2403 & 4.4$\times$10$^{-11}$  &  1.4$\times$10$^{7}$  &  140  &  
     160  &  2.1  &  0.23  &  20.4$\times$11.5  &  0.43   \nl
NGC 4395 & 9.2$\times$10$^{-12}$ & 1.9$\times$10$^{6}$  & 60 & 
     30   &  1.7  &  0.11  &  10.0$\times$8.3   &  1.1    \nl
NGC 6946 & 1.1$\times$10$^{-10}$  &  1.2$\times$10$^{8}$  &  280  & 
     280  &  4.5  &  0.40  &  19.7$\times$16.8  &  0.32   \nl
IC 342  &  1.7$\times$10$^{-10}$  & 2.3$\times$10$^{7}$  &  160  &  
     110  &  2.9  &  0.31  &  13.1$\times$12.8  &  0.47   \nl

\tablecomments{Fluxes, luminosities and surface brightness  
  have been corrected for foreground Galactic extinction based on the
  extinctions listed in Table 1. }

\tablenotetext{a}{Total H$\alpha$ flux. }
\tablenotetext{b}{Total H$\alpha$ luminosity.}
\tablenotetext{c}{Foreground extinction-corrected 
  mean H$\alpha$ surface brightness within the galaxy's
  B = 25 mag arcsec$^{-2}$ isophote (measured from foreground 
  extinction-corrected limiting surface brightness
  of 30 pc cm$^{-6}$ for NGC 6946 and IC 342 and from 15 pc cm$^{-6}$ for
  the rest of galaxies)
  converted to emission
  measure assuming an electron temperature of 10$^4$ K
  (5$\times$10$^{-17}$ erg s$^{-1}$ cm$^{-2}$ arcsec$^{-2}$ 
  corresponds to 25 pc cm$^{-6}$). }
\tablenotetext{d}{Foreground extinction-corrected
  mean H$\alpha$ surface brightness within H$\alpha$ half 
  light radius measured from $-3\sigma$ of background and converted to 
  emission measure.}
\tablenotetext{e}{H$\alpha$ half light radius. }
\tablenotetext{f}{Ratio of the area above a common limiting surface brightness
  that is used to measure $\bar{\Sigma}_{{\tt H}\alpha}$ to the total
  physical area
  within the 25 B mag arcsec$^{-2}$ isophote ($A_{true} = \pi a b / 4$). This
  shows the fractional area used in our measurements of the images. }
\tablenotetext{g}{Major and minor axis diameter for the
  25 B mag arcsec$^{-2}$ isophote (adopted from RC3).}
\tablenotetext{h}{Power-law index of the distribution function for the DIM
  ($\Sigma / \bar{\Sigma} \le 1$) defined as 
  $\frac{dA}{d\lg \Sigma} \propto \left(\frac{\Sigma}{\bar{\Sigma}}\right)^{-n}$. 
  This index  indicates the relative prominence of the DIM in different
  galaxies. A larger  index means a more profuse DIM. }
\enddata

\end{deluxetable}


\clearpage
\begin{figure}
\figurenum{1}
\caption{ Continuum-subtracted H$\alpha$ emission-line images in grayscale
for $a$) M 51, $b$) M 81, $c$) M 101, $d$) NGC 2403, $e$) NGC 4395, $f$) NGC
6946, and $g$) IC 342. North is to the top and
East is to the left. The straight lines mark the slit positions 
we used in the spectroscopic observations (WHL and Wang 1998).
The scale of the images is represented by the length of the slit of 5\arcmin.}
\end{figure}
 
\begin{figure}
\figurenum{2}
\caption{ The growth curve of fractional H$\alpha$ flux
as a function of emission measure cutoff
for our sample galaxies. Each plotted point
corresponds to the relative flux contributed by pixels whose surface brightness
lies between $-3\sigma$ below the background up to the given
surface brightness (emission measure).
Symbols for different objects are explained on the figure.}
\end{figure}
 
\begin{figure}
\figurenum{3}
\caption{ The normalized number of pixels in the H$\alpha$ image within a
given surface brightness interval ($d \lg(\Sigma) = \lg (1.25)$) 
vs. the relative surface
brightness
(scaled by the mean surface brightness in the galaxy $\bar{\Sigma}$). 
The surface brightness plotted starts at minimum values of
30 pc cm$^{-6}$ for NGC 6946 and
IC 342,
and 15 pc cm$^{-6}$ for the other galaxies, 
and it increases by a factor of 1.25 at each
successive step. At each interval of 
surface brightness the number of pixels is counted and normalized by
total number of pixels with surface brightnesses greater than the minimum
values given above. The solid line is the model prediction for high surface
brightness gas based on the typical luminosity function (N(L)$\propto L^{-2}$)
and exponential radial brightness profiles of HII regions.
 See the text for details.}
\end{figure}

\begin{figure}
\figurenum{4}
\caption{ The area used to measure the mean surface brightness $\bar{\Sigma}$
(dark regions)
compared to the total disk area for M 51 and M 101. 
The area and $\bar{\Sigma}$ are used to scale the distribution function
in Figure 3. $a$) M 51. The fraction of total disk area used is 
$\sim$35\%. $b$) M 101. The fraction of total disk area used is 
$\sim$10\%.}
\end{figure}

\begin{figure}
\figurenum{5}
\caption{ The growth of fractional H$\alpha$ flux vs. 
the growth of the fractional covering area. The flux contributed, and the
area occupied, are
measured from the same minimum surface brightnesses as in Figure 3 up
to a given 
relative surface brightness. The values of these relative surface brightnesses
($\Sigma$/$\bar{\Sigma}$)
are marked by ticks increasing from 0.4 (lower-left corner) 
to 3 (upper-right corner) with a stepsize representing
a factor of 1.25 increase in surface brightness. The area is normalized
in the same way as in Fig. 3. See the text for details.}
\end{figure}

\end{document}